\documentclass[aps,prl,twocolumn,showpacs,amsmath,amssymb,floatfix,
superscriptaddress,nofootinbib]{revtex4}
\usepackage{graphicx}

\newcommand {\be}{\begin{equation}}
\newcommand {\ee}{\end{equation}}

\usepackage{xcolor}

\begin{document}
 
 \title{Emergence of slow collective oscillations \\ in neural networks with spike timing dependent plasticity}
 \date{\today}

 \author{Kaare Mikkelsen}
 \email{kbm05@phys.au.dk}
 \affiliation{Dept. of Physics and Astronomy, University of Aarhus,
 Ny Munkegade, Building 1520 - DK-8000 Aarhus C, Denmark}
 
 \author{Alberto Imparato}
 \email{imparato@phys.au.dk}
 \affiliation{Dept. of Physics and Astronomy, University of Aarhus,
 Ny Munkegade, Building 1520 - DK-8000 Aarhus C, Denmark}
 
 \author{Alessandro Torcini}
 \email{alessandro.torcini@cnr.it}
 \affiliation{CNR - Consiglio Nazionale delle Ricerche - Istituto dei Sistemi 
 Complessi, via Madonna del Piano 10, I-50019 Sesto Fiorentino, Italy}
 \affiliation{INFN Sez. Firenze, via Sansone, 1 - I-50019 Sesto Fiorentino, Italy}
 \affiliation{Dept. of Physics and Astronomy, University of Aarhus,
 Ny Munkegade, Building 1520 - DK-8000 Aarhus C, Denmark}

\begin{abstract}
 The collective dynamics of excitatory pulse coupled neurons with spike timing dependent plasticity
 (STDP) is studied. The introduction of STDP induces persistent irregular oscillations between strongly
 and weakly synchronized states, reminiscent of brain activity during slow-wave sleep. We explain the oscillations 
 by a mechanism, the “Sisyphus Effect”, caused by a continuous feedback between the synaptic adjustments and 
 the coherence in the neural firing. Due to this effect, the synaptic weights have oscillating 
 equilibrium values, and this prevents the system from relaxing into a stationary macroscopic state.
\end{abstract}
 
\pacs{05.45.Xt, 87.19.lm, 87.19.lw, 87.19.lj}

\maketitle

Sisyphus was the mythological king of Corinth compelled to roll 
a heavy boulder up a hill, only to watch it roll back down as it approached 
the top. Sisyphus was condemned by Zeus for his iniquity and pride
to repeat eternally his efforts, without any hope of success.
However, in the brain such endless motion can have   
a positive functional relevance. Fluctuating spontaneous activity has been observed in several areas
of the brain \cite{harris}. In particular, irregular oscillations between more and less synchronized 
states have been revealed in the hippocampus during slow-wave sleep and this activity 
has been related to memory consolidation in the neocortex~\cite{buszaki1989}.

Recent studies have suggested synaptic plasticity as a fundamental
ingredient to ensure multistability in neuronal circuits 
\cite{Tass2006BiolCyb,Maistrenko2007Multistability,Lubenov2008, mongillo2012}. 
In particular, spike timing dependent plasticity (STDP) 
is considered one of the central mechanisms
underlying information elaboration and learning in the brain \cite{STDP}.
A series of experiments performed  in vivo and in vitro on neural tissues 
revealed that the strength of a synapse, conveying spikes from a presynaptic 
to a postsynaptic neuron, depends crucially on the precise spike timing of the two
connected neurons~\cite{Markram1997Regulation,Magee1997Synaptically,Bi1998Synaptic}.
The STDP rules prescribe that whenever the presynaptic (postsynaptic) neuron 
fires before the postsynaptic (presynaptic) one, the
synapse is potentiated (depressed). The synapse is modified only if the spikes
occur within certain time intervals ({\it learning windows}). 
Asymmetric learning windows have repeatedly been found experimentally
(e.g. see \cite{Bi2001Synaptic,Bi2002Temporal,FD}).
This asymmetry is a prerequisite, at least in phase
oscillator networks, to observe the coexistence of states characterized 
by different levels of synchrony~\cite{Maistrenko2007Multistability,Tass2006BiolCyb}.
Furthermore, in the presence of propagation delays
STDP can provide a negative feedback mechanism  contrasting highly synchronized
network activity and promoting, in randomly driven networks, 
the emergence of states at the border between randomness and synchrony \cite{Lubenov2008}.

In this Letter a novel deterministic mechanism, the {\it Sisyphus effect} (SE),
able to generate spontaneous fluctuations in a neural network between asynchronous and
synchronous regimes is presented. In particular, we study excitatory pulse coupled 
neural networks with STDP, where the interaction among 
neurons is mediated by $\alpha$-pulses~\cite{Abbott1993Asynchronous}. 
For non plastic interactions, the excitatory coupling leads to synchronization 
only for sufficiently fast synapses~\cite{Vreeswijk1994}.
Furthermore, the desynchronizing effect is amplified at large coupling~\cite{hansel1995}. 
In absence of plasticity the macroscopic activity of the network 
is stationary: asynchronous for large synaptic weights and partially synchronized
for sufficiently weak coupling~\cite{vanVreeswijk1996Partial,hansel1995}. 
  
The introduction of STDP completely modifies the dynamical landscape 
leading to a regime where a strongly and a weakly synchronized state coexist.
The activity of the network is thus characterized by irregular 
oscillations between these two states. These transitions are driven by 
the evolution of the synaptic weights, which in turn is dictated by the 
level of synchrony in the network. For small synaptic weights the system 
is fully synchronized, while above a critical coupling it desynchronizes. 
Furthermore, whenever the network is synchronized (desynchronized)
the synaptic weights tend towards large (small) 
equilibrium values corresponding to asynchronous (synchronous) dynamics.
In summary, the neuronal activity can be represented in terms of an order parameter 
diffusing over an effective free energy landscape displaying two coexisting
equilibrium states. Small (large) synaptic weights tilt the landscape
towards the strongly (weakly) synchronized state, in turn the
induced activity increases (reduces) the weights until
a tilt in the opposite direction occurs. Thus the landscape oscillates endlessly.

{\it The model.} 
We study a fully coupled network of $N$ Leaky Integrate-and-Fire 
neurons, for which the membrane potential $V_i(t) \in [0:1]$ of neuron $i$ 
evolves as:
\begin{equation}\label{eq:V1}
\dot{V}_{i}(t)= a-V_{i}(t)+I_i(t)\, \quad\quad i=1,\cdots, N
\\ \quad ,
\end{equation}
whenever the neuron reaches the 
threshold $V_i=1$, an $\alpha$-pulse $p_\alpha(t) =\alpha^2 t \exp(-\alpha t)$ 
is instantaneously transmitted to all other neurons and $V_i$ is reset to zero.
Furthermore, $a > 1$ is the suprathreshold DC current,  $I_i=g E_i$ the synaptic
current, and $g$ the excitatory {\it homogeneous} coupling.
The field $E_i$ represents the linear superposition of the pulses received
by neuron $i$ and its evolution is ruled by a second order ODE 
(Eq.~(S2) in \cite{epabs}).  For a fully coupled non-plastic network the synaptic weights
associated to the connection from the presynaptic $j$-th neuron to 
the postsynaptic $i$-th one are $w_{ij} = 1$ (apart from the autaptic terms: $w_{ii} =0$).

In the presence of plasticity, we assume that the weights 
evolve in time according to a {\it nearest-neighbour} STDP rule with soft bounds
\cite{STDP,Izhikevich2003BCM,Maistrenko2007Multistability,Chen2010Meanfield,Chen2011}. 
Therefore in the case of a post- (presynaptic) spike, emitted by neuron $i$ ($j$) at time
$t$, the weight $w_{ij}$ is potentiated (depressed) as 
$w_{ij}(t^+) = w_{ij}(t^-)+\Gamma_{ij}(t)$, with
\begin{equation}
\Gamma_{ij}(t) = \left\{\begin{array}{rcl}
p[w_{M}-w_{ij}(t^-)]{\rm e}^{-\frac{\delta_{ij}}{\tau_{+}}} 
& \mbox{if} & \delta_{ij} > 0\\ 
\label{post}
& &\\
- d \, w_{ij}(t^-){\rm e}^{+\frac{\delta_{ij}}{\tau_{-}}} 
& \mbox{if} & \delta_{ij} < 0
\end{array}\right.
\end{equation}
where $\delta_{ij} = t - t^{(j)} > 0$ ($\delta_{ij} = t^{(i)} - t < 0$) is the firing
time difference and $t^{(k)}$ the last firing time of neuron $k$. The potentiation 
and depression factors ($p$ and $d$, resp.) coincide, unless otherwise specified \cite{quant}.
The bounds keep the synapses from achieving unrealistically large values 
or becoming inhibitory, namely $ 0 \le w_{ji} \le w_{M}$.
The learning windows over which post- (pre-) synaptic spikes will cause synaptic 
potentiation (depression) are indicated as $\tau_{+}$ ($\tau_{-}$). 
Following experimental evidences \cite{Bi2001Synaptic}, we assume $\tau_{-} > \tau_{+}$.
The degree of synchronization of the neurons is measured by
the order parameter~\cite{winfree,kuram} 
$R(t)= \left|\frac{1}{N} \sum_k e^{i\theta_k(t)}\right|$,
where  $\theta_k(t)= 2 \pi(t-t^{(k)}_m)/(t^{(k)}_{m+1}-t^{(k)}_m)$
is the phase of the $k$-th neuron at time $t$ between its
$m$-th and $(m+1)$-th spike emission.
A perfectly synchronized (asynchronous) system has $R=1$ ($R=0$),
while intermediate values indicate partial synchronization.

{\it Phase Diagram.}
We analyze how the phase diagram of the network is modified
by the plasticity. In particular, we focus on the variation
of the neuronal coherence by varying the DC current.
Similar results can be obtained by varying the coupling $g$ 
and the pulse width $\alpha$ (as shown in \cite{epabs}). To compare 
with previous results obtained without plasticity, we fix $g=0.4$ and $\alpha=9$ as 
in~\cite{vanVreeswijk1996Partial,zillmer2007}.
In absence of plasticity the homogeneous system exhibits two phases: 
an asynchronous regime with $R \equiv 0$,
and a partially synchronized phase with finite
$R$~\cite{vanVreeswijk1996Partial}. The emergence of
one or the other regime depends crucially on the ratio of two time scales:
the pulse rise time $1/\alpha$ and the interspike-interval (ISI) \cite{Vreeswijk1994,zillmer2007}. 
For slow synapses (relative to the ISI) the system dynamics is
asynchronous, while for sufficiently fast synapses coherent oscillations
emerge. The system becomes fully 
synchronized only for instantaneous synaptic rise times (i.e. $\alpha \to \infty$).
For fixed network size $N$ and pulse shape,
the ISI can be reduced by increasing either the external DC current or 
the synaptic coupling. Therefore partial synchronization is observable for
sufficiently small $a$ or $g$ values (whenever $\alpha \gtrsim 3.4$), 
while incrementing these parameters will desynchronize the system~\cite{hansel1995}
(as shown in Fig. 1a in \cite{epabs}).

\begin{figure}[h]
\begin{center}
\includegraphics*[angle=0,width=7.cm]{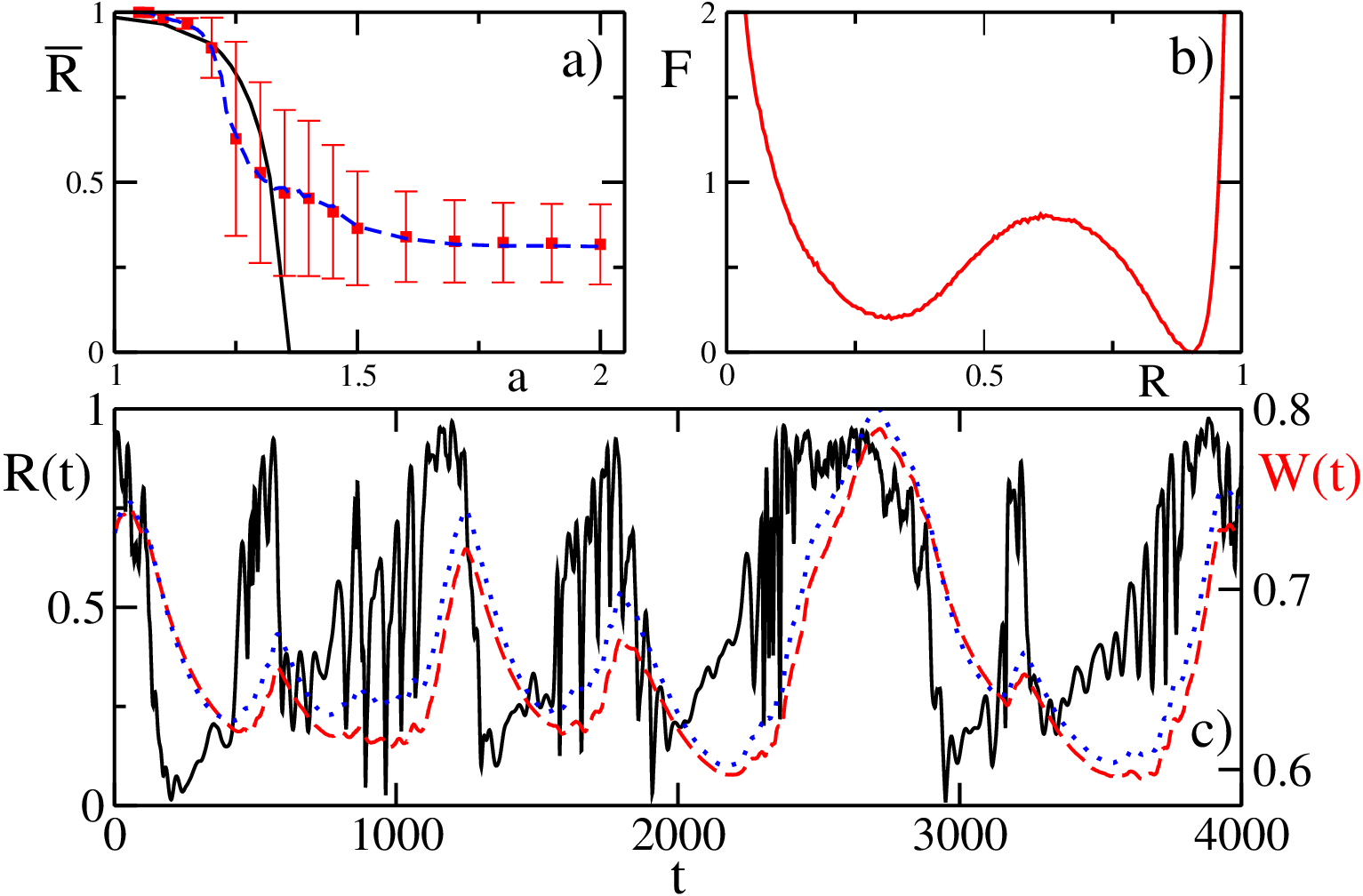}
\end{center}
\caption{(color online) 
(a) Average order parameter $\bar R$ as a function of $a$ for the non
plastic network (black solid line) and in the presence of STDP
for $N=200$ (red filled squares) and $N=500$ (blue dashed line) 
(b) Free energy profile $F(R)$ versus $R$ for $N=200$, obtained
by evaluating $R$ at regular time intervals $\Delta t = 1$ for a time
span $\sim 5 \times 10^6$. (c) Time evolution of 
$R(t)$ (black solid line) and of $W(t)$ (red dashed line) 
for $N=500$. The dotted (blue) line is the $W$ predicted
via Eq.~(\ref{delta}). The data refer to $a=1.3$, $g=0.4$, 
$\alpha=9$, $d=p=0.01$, $\tau_- = 3 \tau_+ = 0.3$, and $w_M=2$,
and are measured after a transient $\sim 10^5$.} 
\label{synch}
\end{figure}

The average level of synchronization $\bar R$ is reported in Fig. \ref{synch}a
as a function of $a$ for the non-plastic and plastic cases. In absence of plasticity 
the system is partially synchronized for low DC currents and asynchronous 
for $a \ge a_c \simeq 1.35$. 
The introduction of plasticity does not alter the scenario at
small $a$-values, where the system is in a {\it high synchronization} (HS) regime.
The main difference is observable in the dynamics
of $R(t)$, which displays irregular oscillations:
the associated Fourier spectrum 
resembles a Lorentzian with a small subsidiary peak around period $\simeq 34-36$.
However,  for sufficiently large currents, namely  $a > 1.5$, 
the asynchronous regime is substituted by a state of {\it low synchronization} (LS)
characterized by a rapidly fluctuating order parameter (over a time scale of the order 
of 70 - 150) with an associated small level of synchronization $\bar R \simeq 0.32 \pm 0.12$.
At intermediate $a$-values, in the range $a \in [1.23;1.46]$, $R$ exhibits 
wide irregular temporal oscillations between values $\simeq 1$ and
zero with characteristic time scales $\simeq 1100 - 1400$.
These latter oscillations represent low frequency fluctuations (LFFs), while rapid fluctuations
are still present over time scales $\simeq 50-60$ (see Fig. \ref{synch}c).

In this Letter we will mainly focus on the intermediate regime, 
fixing $a=1.30$, 
where the LFF of $R(t)$ resembles the evolution of a particle in a 
double well potential subject to thermal fluctuations. To 
clarify this analogy we have estimated the PDF, $P(R)$, of the order parameter,
by examining  its trajectory for a sufficiently long time span,
and derived the associated {\it free energy} profile as $F(R)= - \log P(R)$.
As shown in Fig. \ref{synch}b, $F(R)$ exhibits two minima 
corresponding to a HS phase at $R_H \simeq 0.905$ and a LS 
state at $R_L \simeq 0.32$. The 2 coexisting minima are
separated by a saddle, located at $ R_S \simeq 0.61 $.
As clarified in the following the jumps between minima are driven 
by the macroscopic evolution of network plasticity.
The rapid fluctuations, present in all regimes, 
are instead due to the microsopic evolution of the synaptic weights, which can be 
interpreted as a noise source for the dynamics of $R(t)$. The analysis of these
{\it noise-induced oscillations} goes beyond the scope of this Letter
and it is left for future studies.

{\it Constrained phase diagram} 
As shown in Fig.~\ref{synch}c the LFFs of $R(t)$ are associated with oscillations in 
the average synaptic weight $W(t)\equiv \sum_{i,j} w_{ji}(t)/N(N-1)$.
In particular when the system is in a HS (LS) state $W$ increases (decreases). 

\begin{figure}[h]
\begin{center}
\includegraphics*[angle=0,width=6.4cm]{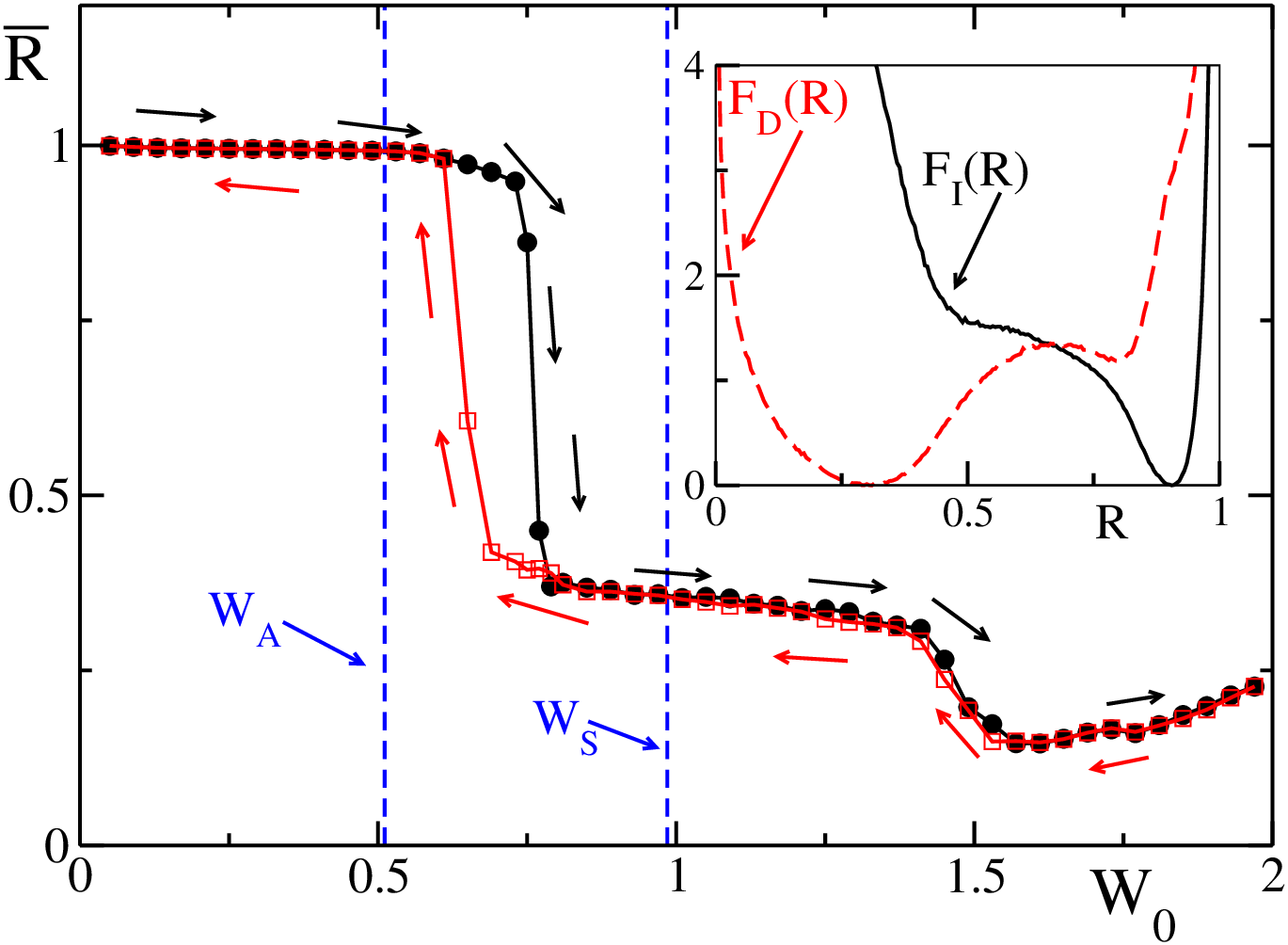}
\end{center}
\caption{(color online) 
$\bar R$ versus $W_0$ as measured 
for increasing (black filled circles) and decreasing
(empty red squares) $W_0$.
The (blue) vertical dashed lines indicate the 
fixed point values $W_{S}$ and $W_{A}$~\cite{epabs}.
Results averaged over 8 different initial conditions, $T_S = 1,000$,
$\Delta W_0 = 0.02 $ (for clarity only one point every two is shown).
(Inset) Conditional free energy profiles $F_I(R)$ (black solid line)
and $F_D(R)$ (red dashed line) obtained during USs. Both curves are vertically shifted to achieve zero
as minimal value. Parameters as in Fig. \ref{synch} and $N=200$.
} 
\label{R_W_locked}
\end{figure}

To better investigate the origin of these correlations and the
interaction between the STDP induced synaptic dynamics 
and the level of synchronization in the system we perform the following 
numerical experiments. We simulate the system by constraining the
synaptic weights to have a constant average value $W_0$, by rescaling, at regular time intervals, the weights $w_{ij}$. 
Initially, $W_0=0$ and we follow the evolution of the system for a time span $T_S$.
We then perform a new simulation for the same time lapse with a larger $W_0$ value, starting 
from the last configuration of the previous run. The procedure is repeated by increasing $W_0$ 
at regular steps $\Delta W_0$ until $W_0 =w_{M}$ is reached. Then with the same protocol 
$W_0$ is decreased (in steps of $\Delta W_0$) until finally $W_0$ returns to zero \cite{note3}.
The results of these simulations are shown in Fig. \ref{R_W_locked} for $N=200$.
At low $W_0$ the system is fully synchronized,
while with increasing $W_0$ the system desynchronizes via a
discontinuous transition. By further increasing $W_0$ the level of synchronization continues to decrease
and another smooth transition seems to occur.
For the explanation of the SE it is sufficient to limit
the analysis to the first transition.

As shown in Fig. \ref{R_W_locked}, the constrained system exhibits a hysteretic transition 
from HS to LS (from LS to HS) for $W_0^{(1)} = 0.76(5)$ ($W_0^{(2)} = 0.65(5)$)
by increasing (decreasing) the control parameter $W_0$ \cite{note2}. 
This implies that in the interval $[W_0^{(2)},W_0^{(1)}]$ the two regimes coexist
and that HS or LS is observable depending on the initial state of the network.

{\it Mean field synaptic evolution.} In order to gain some insight into the 
evolution of the system during unconstrained simulations (USs), let us
consider a mean field equation for the synaptic
weight evolution. The average synaptic weight modification $\Gamma$, for
each presynaptic spike, can be written as~\cite{Izhikevich2003BCM}
\begin{equation}
\Gamma(t)= p(w_{M}- W) \int_0^\infty d\delta P(\delta) 
{\rm e}^{\frac{-\delta}{\tau_{+}}} 
- d  W   \int_{-\infty}^0 d\delta P(-\delta) 
{\rm e}^{\frac{\delta}{\tau_{-}}}
\label{delta}
\end{equation}
where $P(\delta)$ is the PDF of the time differences $\delta$ 
between postsynaptic and presynaptic firing measured. 
To test the predictive value of Eq.~(\ref{delta}), we have measured from an US
$P(\delta)$ at regular intervals $\Delta t$. By employing this information
we can predict quite well the evolution of the synaptic weight as 
$W(t+\Delta t) = W(t) + \Gamma(t)$ (see Fig. \ref{synch}c).
 
By assuming that the postsynaptic neuron is firing with 
period $T_0$, we are able to derive the time difference distribution $P(\delta)$
for the two limiting cases: fully synchronized and asynchronous dynamics.
In the fully synchronized (asynchronous) situations we expect 
a distribution of the form $P_S(\delta)= {\cal D}(\delta) +
{\cal D}(\delta-T_0)$
($P_A(\delta)= 1/T_0$) defined in the interval $[0:T_0]$. Here
${\cal D}$ denotes a Dirac delta function.	
These guesses are essentially confirmed by direct 
USs as shown in Fig. 3 in \cite{epabs}.
Therefore in these two cases an analytical estimation of 
$\Gamma$ can be obtained. Furthermore, 
in both cases $\Gamma$ vanishes for a finite value
of the average synaptic weight, namely $W_S$ ($W_A$) for the synchronized
(asynchronous) situation. Furthermore, for $ W < W_S$ ($W > W_S$)
the synapses are in average potentiated (depressed) and analogously
for the asynchronous case. This implies that  $W_S$ ($W_A$)
is a stable attractive point for the dynamics of $W$ in the synchronized
(asynchronous) regime (for a definition of $W_S$ and $W_A$
see Eq.~(S8) and (S10) in \cite{epabs}).

{\it Sisyphus mechanism.} We are now able to explain the
behavior reported in Fig. \ref{synch}c for $R(t)$ and $W(t)$.
Let us suppose that the system is in the HS phase
with an associated low coupling $W < W_0^{(1)}$. However, in this situation the attractive 
fixed point $W_S$ is above the transition point $W_0^{(1)}$ (see Fig. \ref{R_W_locked}). 
Therefore $W$ keeps increasing, until for $W > W_0^{(1)}$ the system 
starts to desynchronize and to approach the LS state.
In this phase the $P(\delta)$ becomes 
almost flat (see  Fig. 3b in \cite{epabs}) and the attractive point for the synaptic 
evolution will be $W_A$, located below $W_0^{(2)}$. The motion towards 
$W_A$ leads to a decrease of $W$. Whenever the average synaptic weight
crosses $W_0^{(2)}$ the neurons begins to resynchronize. Finally,
the system will return to the HS state from where it started. The cycle will repeat indefinitely 
and is the essence of the SE.

The above arguments are approximate, because the system is never exactly fully
synchronized or desynchronized, instead it passes through a continuum of states,
each associated to a different fixed point in $W$-space.
The relevant aspect is that the fixed points associated to the HS (LS) phase
are larger than the transition point $W_0^{(1)}$ (smaller than  $W_0^{(2)}$). 
As we have verified this is indeed the case, therefore the mechanism
is still valid. To perform a direct test of the validity of our analysis, we have 
measured the PDF of $R$ conditioned to the fact that 
$W$ was increasing (decreasing) during an US. From these PDFs 
we derived the corresponding free energy profile $F_I(R)$ ($F_D(R)$). 
As shown in the inset of Fig. \ref{R_W_locked} $F_I$ has a unique
minimum at $R_H$, while $F_D$ has an absolute minimum at $R_L$ and a shoulder
around $R \simeq 0.8$. These results confirm that the equilibrium attractive values 
for $W$ are located opposite to the transition points, because when the system is in 
the HS (LS) regime the synaptic weights increase  (decrease) continuously
trying to reach the corresponding fixed points.

\begin{figure}[h]
\begin{center}
\includegraphics*[angle=0,width=6.4cm]{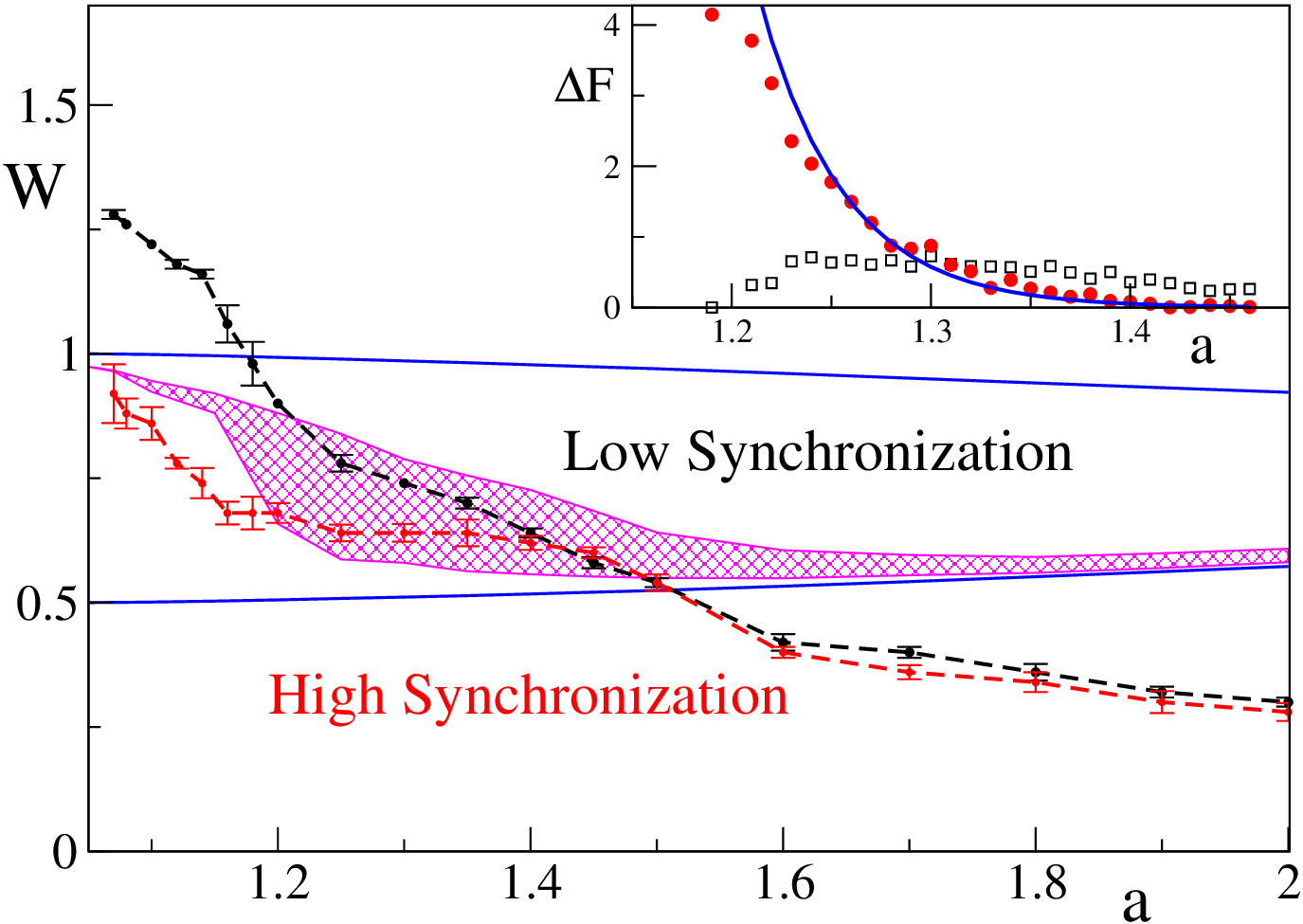}
\end{center}
\caption{(color online) 
$W$ versus the DC current $a$. The shaded area represents
the $W$-values measured during USs. The upper black (lower red) 
dashed line refers to the estimated $W_0^{(1)}$ ($W_0^{(2)}$). 
The error bars have been evaluated over
5 different realizations of the constrained simulations.
The upper (lower) solid blue line represents the fixed point
values $W_S$ ($W_A$). (Inset) The red circles 
(black squares) refer to the free energy barrier $\Delta F$ 
separating the HS (LS) state
from the saddle. The blue line is an exponential fit to the
HS barrier height. The remaining parameters 
as in Fig. \ref{synch} and $N=500$
} 
\label{W_a}
\end{figure}

The SE should be active whenever the transition values $W_0^{(1)}$ and $W_0^{(2)}$ 
are both contained within the interval $[W_A, W_S]$. To verify this statement
we have measured $W_0^{(1)}$, $W_0^{(2)}$  and the fixed points
for various DC currents within the interval $0 < a \le 2$ 
(data shown in Fig. \ref{W_a}). 
We observe that the transition is hysteretic 
in the interval $a \in ]0;1.40]$, while for larger values 
$W_0^{(1)}$ and $W_0^{(2)}$ essentially coincide. Furthermore,
$W_0^{(1)}$ becomes larger than $W_S$ at $a \simeq 1.18$, while
$W_A \ge W_0^{(1)},W_0^{(2)}$ for $a \ge 1.50$. Thus we expect that
$F(R)$ exhibits two coexisting minima, due to the SE,
when $ 1.18 \le a \le 1.50$. To verify this conjecture
we estimate the free energy barrier heights $\Delta F$ separating 
the HS and the LS state from the intermediate saddle for various $a$ values.
As shown in the inset of Fig. \ref{W_a}, the barrier associated to 
the HS state diverges exponentially when approaching $a \simeq 1.18$. 
Therefore, the HS regime is the only possible at smaller $a$-values.
On the other hand the two minima merge and the associated barriers
vanish for $a \ge 1.48$ indicating that the LS state is the unique
remaining at large $a$.  Furthermore, the distributions of the $W$-values measured 
during USs are reported in Fig.~\ref{W_a} as a shaded area:
these values include the transition interval $[W_0^{(1)};W_0^{(2)}]$ for $ 1.20 \le a \le 1.48$. 

In conclusion, the SE should be observable 
in pulse coupled neural networks whenever the excitation has a desynchronizing effect. 
This is in general verified for any kind of neuronal response (type I or type II) 
for sufficiently slow synaptic interactions~\cite{Vreeswijk1994,hansel1995}.
Furthermore, we have verified that the SE persists by setting $p > d$, as suggested by 
experimental evidences~\cite{FD}.

\begin{acknowledgments}
AT acknowledges the VELUX Visiting Professor Programme 2011/12 
and the Aarhus Universitets Forskningsfond for the support received during his stays 
at the University of Aarhus (Denmark).
This work is part of the activity of the Marie Curie Initial  Training Network 'NETT'
project \# 289146 financed by the European Commission.
We thank S. Lepri and S. Luccioli for useful discussions
as well as for a careful reading of this Letter prior to the submission.
\end{acknowledgments}

\vspace{2.0 cm}

\centerline{\bf ADDITIONAL MATERIAL}

\section{The Model}

We study a fully coupled network of $N$ Leaky Integrate-and-Fire (LIF)
neurons, for which the membrane potential $V_i(t) \in [0:1]$ of neuron $i$ 
evolves as:
\begin{equation}\label{eq:V1}
\dot{V}_{i}(t)= a-V_{i}(t)+I_i(t)\, \quad\quad i=1,\cdots, N
\\ \quad ,
\end{equation}
where $a > 1$ is the suprathreshold DC current, and $I_i$ the synaptic
current due to the coupling with the rest of the network. 
Following \cite{abbott_vanvreeswik}, we assume that
whenever the $i$-th neuron reaches the
threshold $V_i=1$, an $\alpha$-pulse $p_\alpha(t) =\alpha^2 t \exp(-\alpha t)$ is instantaneously 
transmitted to all the other neurons and its membrane potential is reset to $V_i=0$. 
The synaptic current can be written as $I_i(t) = g E_i(t)$, with $g  >  0$ representing
the excitatory {\it homogeneous} coupling while the field $E_i(t)$ is
given by the linear superposition of the pulses received
by neuron $i$ in the past. For $\alpha$-pulses
the time evolution of the field $E_i(t)$ is ruled by
the following second order differential equation:
\begin{equation}
\label{eq:E}
\ddot E_i(t) +2\alpha\dot E_i(t)+\alpha^2 E_i(t)= 
\frac{\alpha^2}{N-1}\sum_{n|t_n< t} w_{ij} \delta(t-t_n) \ .
\end{equation}
with $\{t_n\}$ being the firing times until the present time $t$.
For a fully coupled network, in absence of plasticity,
the synaptic weights $w_{ij}$ appearing in Eq. (\ref{eq:E})
are all equal to one (apart from the autaptic terms which are set to zero).
In presence of plasticity, we assume that the synaptic weights
evolve in time according to a STDP rule with soft bounds, namely
\cite{STDP,Maistrenko2007Multistability,Chen2010Meanfield}

\begin{equation}\label{eq:dynW}
\dot{w}_{ij}(t)= p (w_{max}-w_{ij}(t)) A_j S_i - d w_{ij}(t) B_i S_j 
\quad ,
\end{equation}

where $d$ (resp. $p$) is the potentiation (resp. depression) amplitude, and $S_k$ 
represents the series of spiking times of neuron $k$ until time $t$. 
The presence of the bounds implies that $ 0 \le
w_{ij} \le w_{max}$. By assuming that the synapses have memory just of 
the last emitted spike ({\it nearest-neighbour} STDP rule), the variables 
$A_j$ (resp. $B_i$) evolves as follows \cite{STDP,Chen2010Meanfield}
\begin{equation}\label{eq:prepost}
\tau_{+} \dot{A_j} = - A_j +(1-A_j) S_j \enskip
, \enskip
\tau_{-} \dot{B_i} = - B_i +(1-B_i) S_i 
\end{equation}
where $\tau_{+}$ (resp. $\tau_{-}$) are the time windows over
which post- (resp. pre-) synaptic spikes will cause potentiation
(resp. depression) of the synapse. As pointed out by Izhikevich and Desai~\cite{Izhikevich2003BCM} 
the nearest neighbour implementation of the STDP rule is the only one consistent with the classical
results on long-term potentiation and depression by
Bienenstock-Cooper-Munro~\cite{BCM}

Therefore in the case of a post-synaptic (pre-synaptic)
spike, emitted by neuron $i$ ($j$) at time
$t$, the weight $w_{ij}$ is potentiated (depressed) as 
$w_{ij}(t^+) = w_{ij}(t^-)+\Gamma_{ij}(t)$, with
\begin{equation}
\Gamma_{ij}(t) = \left\{\begin{array}{rcl}
p[w_{M}-w_{ij}(t^-)]{\rm e}^{-\frac{\delta_{ij}}{\tau_{+}}} 
& \mbox{if} & \delta_{ij} > 0\\ 
\label{post}
& &\\
- d \, w_{ij}(t^-){\rm e}^{+\frac{\delta_{ij}}{\tau_{-}}} 
& \mbox{if} & \delta_{ij} < 0
\end{array}\right.
\end{equation}
where $\delta_{ij} = t - t^{(j)} > 0$ ($\delta_{ij} = t^{(i)} - t < 0$) is the firing
time difference and $t^{(k)}$ the last firing time of neuron $k$.

Since the plasticity rule depends critically on the precision of the spiking
events, it is necessary to employ an accurate integration scheme to update
the evolution equations. For this purpose we adopted an event-driven algorithm 
with event queue conjugating high accuracy with a fast implementation~\cite{zillmer2007,Chen2011}.

The degree of synchronization of the neuronal population can be characterized in terms
of the order parameter~\cite{winfree,kuram}
\begin{equation}
R(t)=  \left|\frac{1}{N} \sum_k e^{i\theta_k(t)}\right| \quad ,
\label{kura}
\end{equation}
where 
\begin{equation}
 \theta_k(t)=2\pi \frac{(t-t^{(k)}_m)}{(t^{(k)}_{m+1}-t^{(k)}_m)}
\label{angle}
\end{equation}
is the phase of the $k$-th neuron at time $t$ between its 
$m$-th and $m+1$-th spike emission, occurring at times
$t^{(k)}_m$ and $t^{(k)}_{m+1}$, respectively.
A non zero $R$ value is an indication of partial
synchronization, perfect synchronization is
achieved for $R=1$, while a vanishingly small $R \sim 1/\sqrt{N}$ is 
observable for asynchronous states in finite systems.

\section{Phase Diagram}

\begin{figure}[h]
\begin{center}
\includegraphics*[angle=0,width=7.5cm]{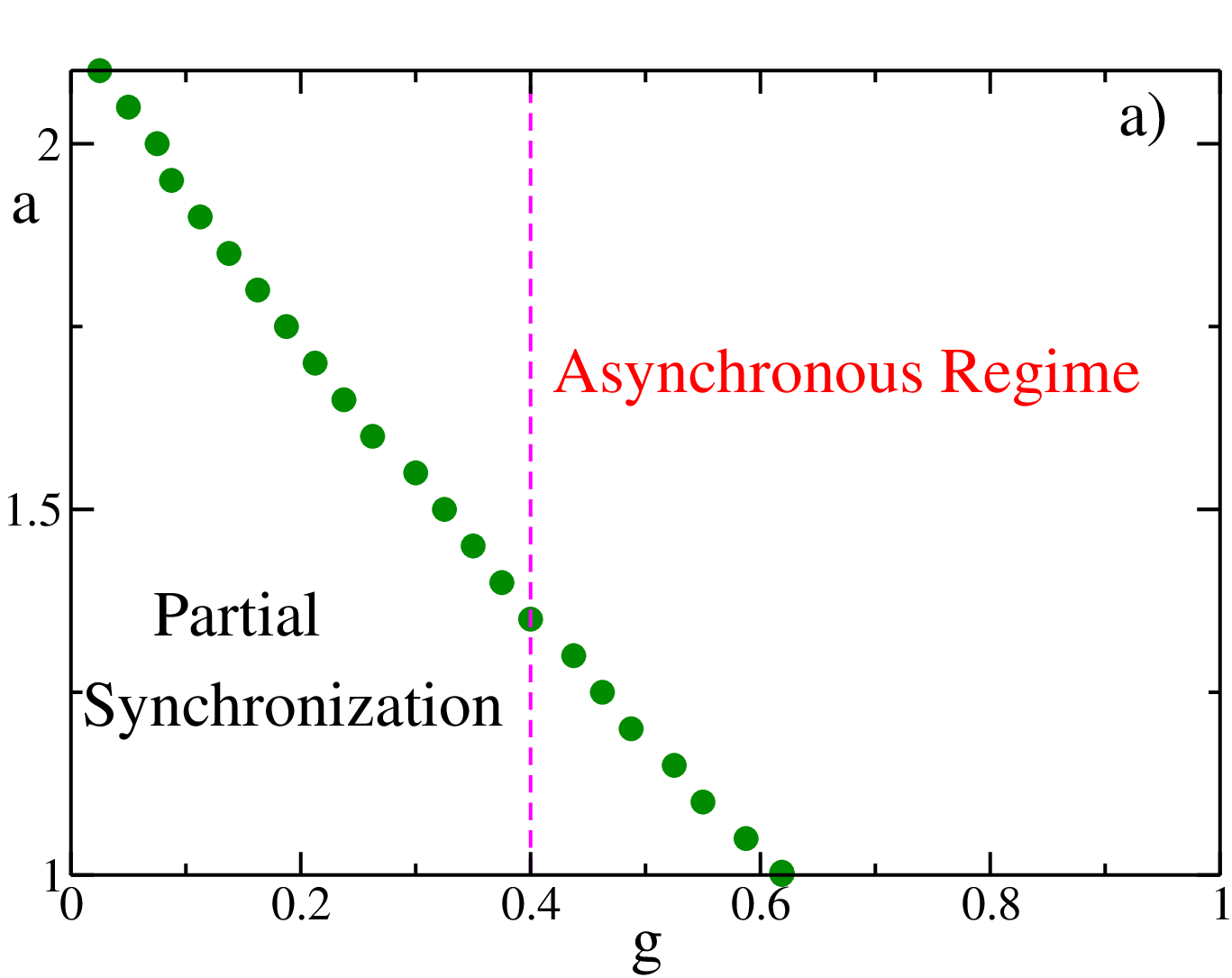}
\includegraphics*[angle=0,width=7.5cm]{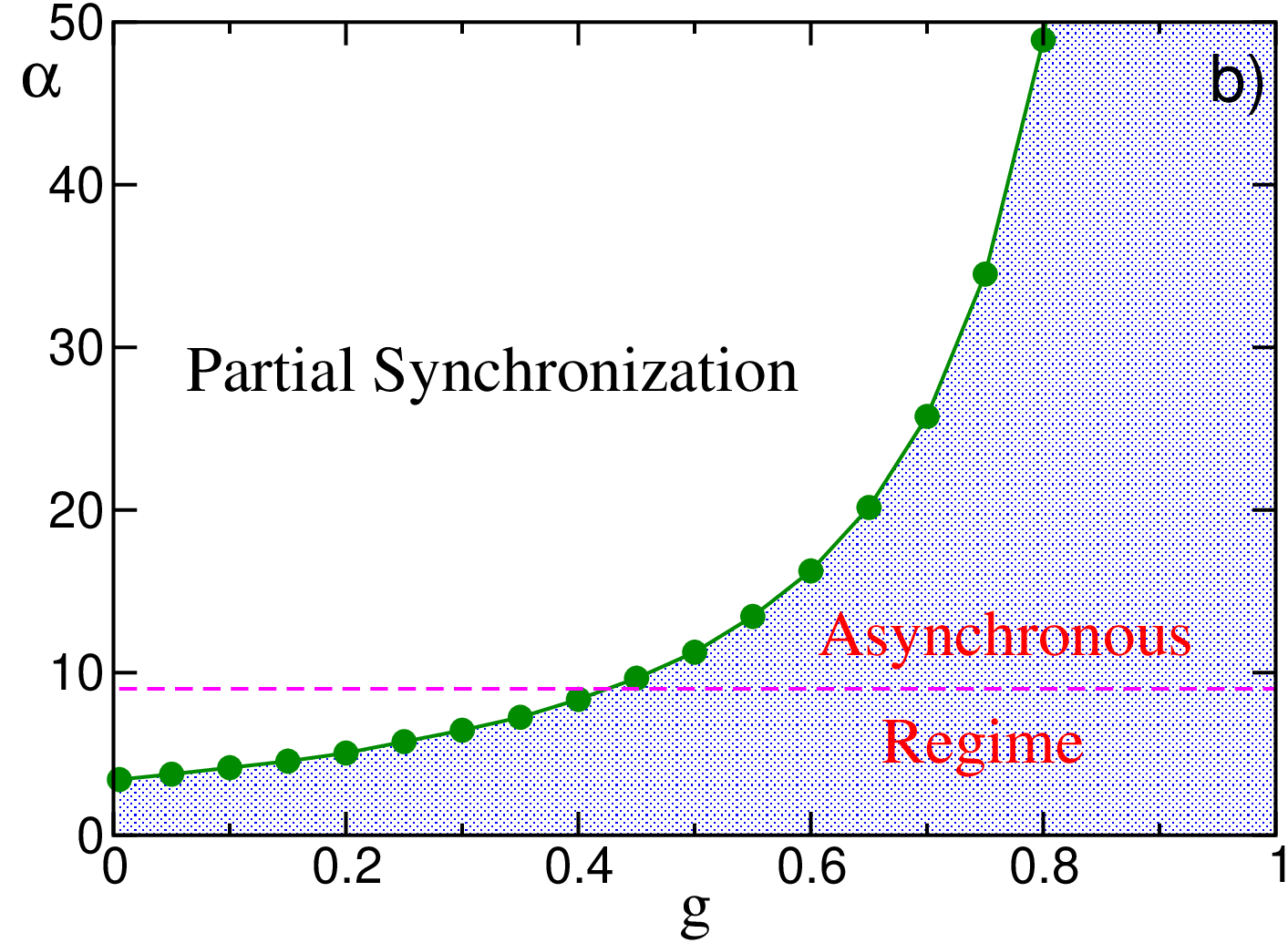}
\end{center}
\caption{(color online) 
Phase diagram for the homogeneous model without plasticity
in (a) we report the phase plane $(g,a)$, 
while in (b) the phases are shown in the $(g,\alpha)$-plane.
The (green) filled circles indicate the critical values for which the 
asynchronous state (namely, the splay state) becomes unstable;
the dashed (magenta) line refers to $g=0.4$ in (a) and $\alpha=9$ in (b).
The other parameters are fixed to $\alpha=9$ in (a) and $g=0.4$ in (b) and 
the system size is  $N=100$.
} 
\label{phasediag}
\end{figure}

We report in Fig. \ref{phasediag} the phase diagram for the non plastic fully connected network for 
$N=100$ in the planes $(g,a)$ and $(g,\alpha)$.
In this case the asynchronous regime corresponds to a 
so-called {\it splay state} \cite{abbott_vanvreeswik,zillmer2007}:
this is an exact solution for the system which is perfectly asynchronous ($R\equiv0$).
For this periodic solution we are able to perform an analytical stability analysis \cite{zillmer2007} ,
reported in Fig. \ref{phasediag}. The (green) circles in Fig.~\ref{phasediag} indicates the marginal 
stability line for the the splay state, it is known that whenever the splays state looses its stability
it gives rise to a partially synchronized regime via a Hopf supercritical bifurcation
~\cite{abbott_vanvreeswik}. With reference to the parameter values considered in this Letter, 
partial synchronization emerges for $a \le a_c \simeq 1.35 $ for fixed coupling $g=0.4$,
and for $g \le g_c \simeq 0.4676 $ for fixed pulse width $\alpha=9$.

In order to characterize the dynamical phases observed in the plastic
and non plastic networks, we have also estimated the order parameter $\bar R$
averaged over a certain time span. 

\begin{figure}[h]
\begin{center}
\includegraphics*[angle=0,width=7.5cm]{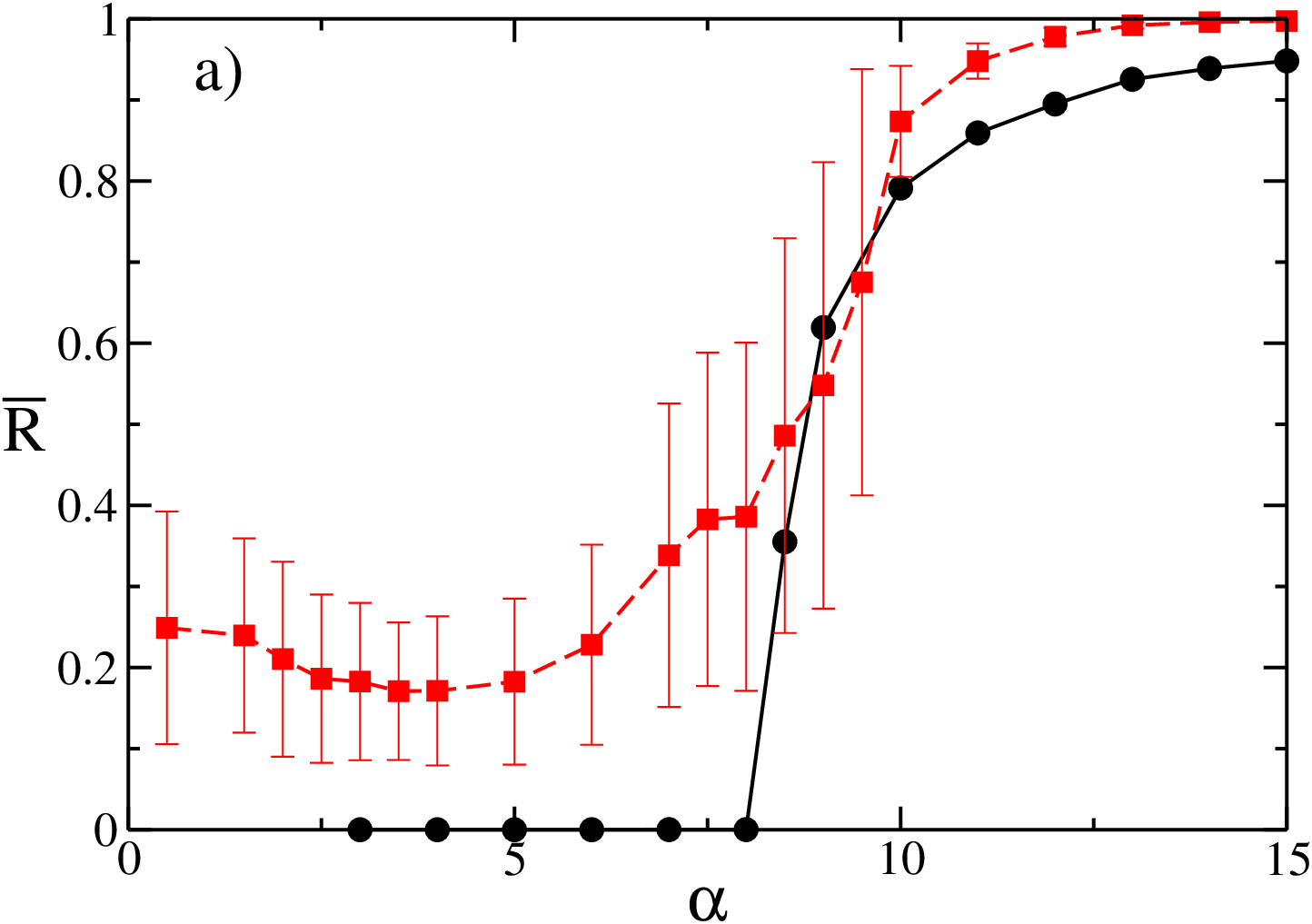}
\includegraphics*[angle=0,width=7.5cm]{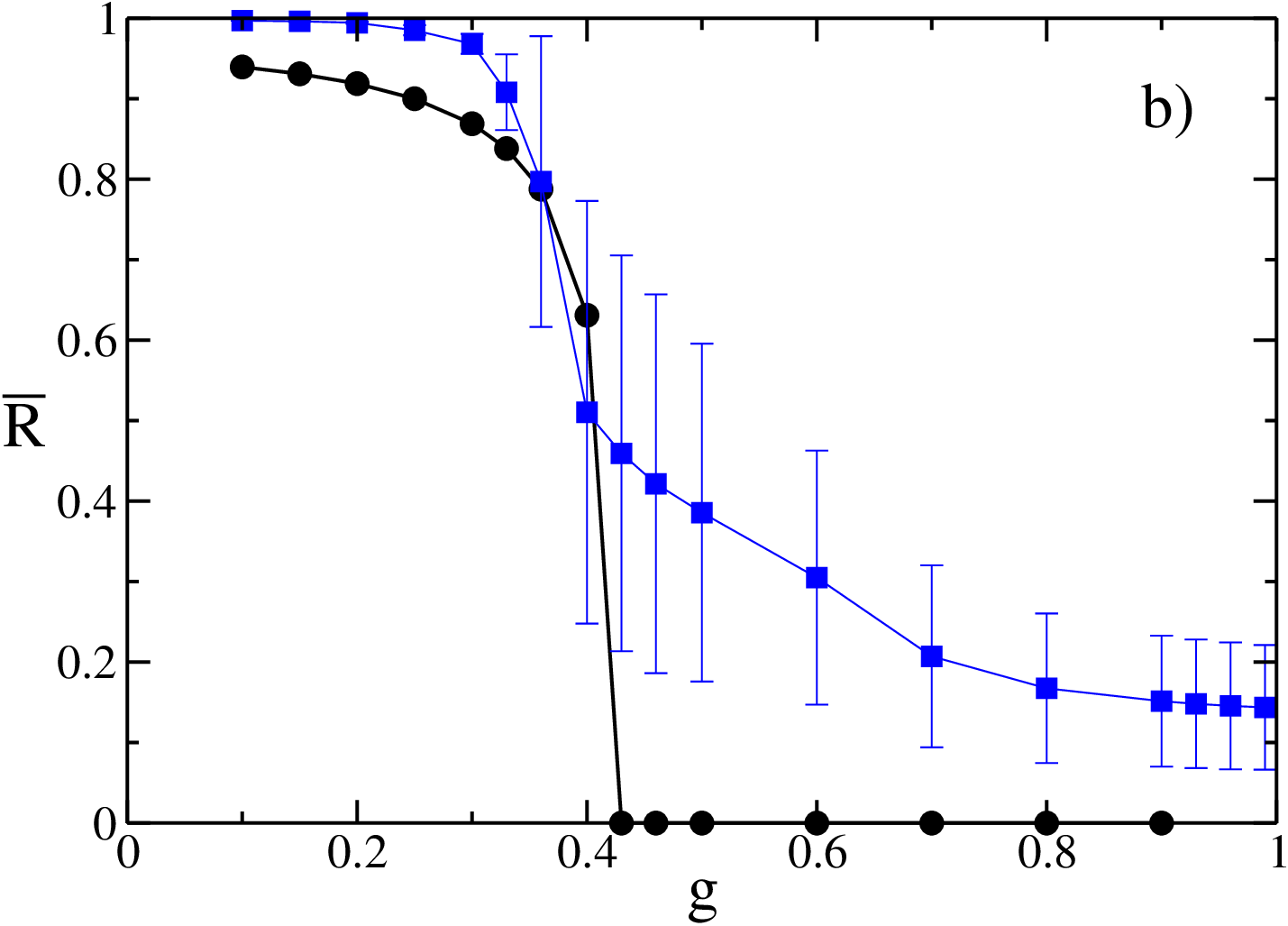}
\end{center}
\caption{(color online) 
(a) Average order parameter $\bar R$ versus $\alpha$ for  the systems
without plasticity (black circles) and with STDP (red squares);
(b) $\bar R$ versus $g$ for  the systems
without plasticity (black circles) and with STDP (blue squares).
The parameters are $a=1.3$ and $N=200$, in (a) $g=0.4$
and in (b) $\alpha=9$. The average
have been performed over time spans $\simeq 10^4 - 10^5$ time units,
after discarding transients of duration $\simeq 10^4 - 10^5$. 
} 
\label{magnetization}
\end{figure}

The behaviour of $\bar R$ as a function of $g$ and $\alpha$ are reported in
in Fig. \ref{magnetization}, similarly to the non plastic case the system desynchronizes
also in presence of STDP for increasing synaptic coupling $g$ and pulse
rise times $1/\alpha$. However, the perfectly asynchronous regime is
substituted by a low synchronization phase where $\bar R \sim 0.2$.
Furthermore in proximity of the transition region from high synchronous to the
low synchronous regime $R(t)$ exhibits large low frequency fluctuations
as in the case studied in the Letter.

\noindent

\section{Time difference distributions}

\noindent
The probability density distributions (PDF) for the time differences
$\delta$ between postsynaptic and presynaptic firing time can be
easily derived in two limiting case by assuming a constant inter-spike
interval (ISI) $T_0$ for the post-synaptic neuron. For fully synchronized neurons, we expect 
the pre- and postsynaptic neurons to fire
together $\delta = 0$ or with a delay given by the ISI, namely $\delta = T_0$.
Therefore the distribution will be $P_S(\delta)= {\cal D}(\delta) +
{\cal D}(\delta-T_0)$, where ${\cal D}(x-x_0)$ denotes a Dirac delta function centered
in $x_0$.
The other situation we consider is that corresponding to perfect asynchrony,
in this case we expect that $\delta$ will take all values in the interval $[0:T_0]$
and all the values will be equiprobable therefore we expect $P_A(\delta)= 1/T_0$.

\begin{figure}[h]
\begin{center}
\includegraphics*[angle=0,width=7.5cm]{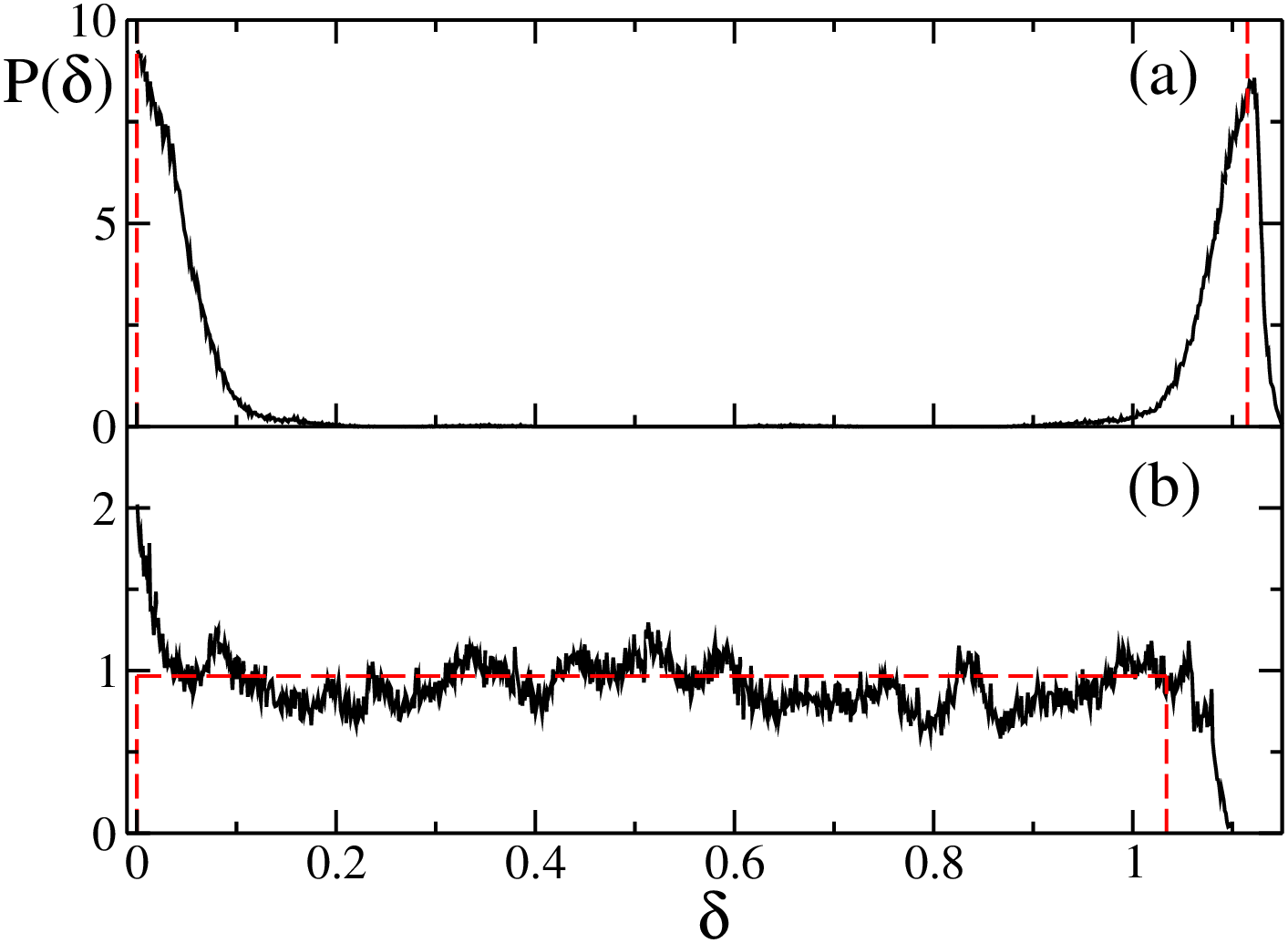}
\end{center}
\caption{(color online) 
Probability distribution functions $P(\delta)$ as obtained during an unconstrained simulation 
by considering the $N \times (N-1)$ $\delta_{ij}$ values associated to the last $N$ spikes
preceeding a strongly (resp. weakly) synchronized state corresponding to
an order parameter value $R \simeq 0.98$ (resp. $R \simeq 0.11$). 
The estimated PDF are reported in panel (a) and (b), respectively. 
The red dashed lines refer to $P_S$ in (a) and $P_A$ in (b) defined in the text.
The parameters of the simulation are $\alpha=9$, $\tau_- = 3 \tau_+ = 0.30$, $d=p=0.01$,
$g=0.4$ and the network size is $N=500$
} 
\label{distribution}
\end{figure}

In Fig. \ref{distribution} we compare the predictions with the measured
distributions, the agreement is reasonable in view of the fact that
the two considered states do not correspond exactly to $R=1$ and $R=0$ and that
the ISI is not constant over all the neuronal population. Therefore,
we can consider $P_A(\delta)$ and $P_S(\delta)$ as reasonable approximation of the true distributions
in the two extreme cases achieved by the system during its evolution.

\section{Mean field synaptic evolution}

By assuming that the post-synaptic neuron fires with constant period $T_0$ we
can perform the integrals appearing in Eq. (3) in the letter in the 2 limiting cases discussed above.

\subsection{Asynchronous dynamics}

In this situation $P(\delta) = P_A (\delta)$ and we can rewrite Eq. (3) as follows
\begin{equation}
\Gamma_A =  \frac{p \tau_+}{T_0} \left[ (w_{M}- W)(1- {\rm e}^{-\frac{T_0}{\tau_{+}}}) 
- 3  W   (1- {\rm e}^{-\frac{T_0}{\tau_{-}}}) \right]
\label{PA}
\end{equation}
where we have assumed $p=d$ and $\tau_- = 3 \tau_+$ as for the most part of the simulations
studied in the Letter. The quantity $\Gamma_A$ vanishes for $W=W_A$ and it is positive
(resp. negative) for $W < W_A$ (resp. $ W > W_A$), therefore for the dynamics of $W(t)$
\begin{equation}
W_A = \frac{w_{M} \left(1- {\rm e}^{-\frac{T_0}{\tau_{+}}}\right)}{4-{\rm e}^{-\frac{T_0}{\tau_{+}}}-3{\rm e}^{-\frac{T_0}{\tau_{-}}}}
 \label{WA}
\end{equation}
is a stable fixed point.

\subsection{Fully synchronized dynamics}
 
For the fully synchronized situation $P(\delta) = P_S (\delta)$ and Eq. (3) becomes
\begin{equation}
\Gamma_S =  p  \left[ w_{M}(1+{\rm e}^{-\frac{T_0}{\tau_{+}}}) 
- W   (2 +  {\rm e}^{-\frac{T_0}{\tau_{+}}}       + {\rm e}^{-\frac{T_0}{\tau_{-}}}) \right]
\label{PS}
\end{equation}
where we have assumed once more $p=d$ and $\tau_- = 3 \tau_+$. 
The quantity $\Gamma_S$ vanishes for $W=W_S$ and it is positive
(resp. negative) for $W < W_S$ (resp. $ W > W_S$), therefore the solution
\begin{equation}
W_S = \frac{w_{M} \left(1+{\rm e}^{-\frac{T_0}{\tau_{+}}}\right)}{2+{\rm e}^{-\frac{T_0}{\tau_{+}}}+{\rm e}^{-\frac{T_0}{\tau_{-}}}}
 \label{WS}
\end{equation}
represents an attractive fixed point.


\begin{thebibliography}{10}
\bibitem{quant}
 All the quantities appearing in the model are dimensionless.
\bibitem{note3}
For each $W_0$, $\bar R$ is estimated only over 
the second half of the simulation, thus discarding a transient $T_S/2$  for each run.
\end{thebibliography}
\end{document}